\documentclass{ieeeaccess}
\usepackage{amsmath,amssymb,amsfonts}
\usepackage{algorithmic}
\usepackage{graphicx}
\usepackage{textcomp}
\usepackage{multirow}
\usepackage{caption}

\def\BibTeX{{\rm B\kern-.05em{\sc i\kern-.025em b}\kern-.08em
    T\kern-.1667em\lower.7ex\hbox{E}\kern-.125emX}}
\begin{document}
\history{Date of publication xxxx 00, 0000, date of current version xxxx 00, 0000.}
\doi{10.1109/ACCESS.2017.DOI}

\title{Scientific Paper Recommendation: A Survey}
\author{\uppercase{Xiaomei Bai}\authorrefmark{1},
\uppercase{Mengyang Wang\authorrefmark{2}, Ivan Lee\authorrefmark{3}, \IEEEmembership{Senior Member, IEEE}, Zhuo Yang\authorrefmark{2}, Xiangjie Kong\authorrefmark{2}, \IEEEmembership{Senior Member, IEEE} and Feng Xia
}\authorrefmark{2}, \IEEEmembership{SENIOR MEMBER, IEEE}}
\address[1]{Computing Center, Anshan Normal University, Anshan 114007, China}
\address[2]{School of Software, Dalian University of Technology, Dalian 116620, China}
\address[3]{School of Information Technology and Mathematical Sciences, University of South Australia, Adelaide SA 5001, Australia}
\tfootnote{This work was partially supported by the Fund for Promoting the Reform of Higher Education by Using Big Data Technology, Energizing Teachers and Students to Explore the Future (2017A01002), Liaoning Provincial Key R\&D Guidance Project (2018104021) and Liaoning Provincial Natural Fund Guidance Plan (20180550011).}


\corresp{Corresponding author: Xiaomei Bai (e-mail: xiaomeibai@outlook.com).}

\begin{abstract}
Globally, recommendation services have become important due to the fact that they support e-commerce applications and different research communities. Recommender systems have a large number of applications in many fields including economic, education, and scientific research. Different empirical studies have shown that recommender systems are more effective and reliable than keyword-based search engines for extracting useful knowledge from massive amounts of data. The problem of recommending similar scientific articles in scientific community is called scientific paper recommendation. Scientific paper recommendation aims to recommend new articles or classical articles that match researchers' interests. It has become an attractive area of study since the number of scholarly papers increases exponentially. In this survey, we first introduce the importance and advantages of paper recommender systems. Second, we review the recommendation algorithms and methods, such as Content-Based methods, Collaborative Filtering methods, Graph-Based methods and Hybrid methods. Then, we introduce the evaluation methods of different recommender systems. Finally, we summarize open issues in the paper recommender systems, including cold start, sparsity, scalability, privacy, serendipity and unified scholarly data standards. The purpose of this survey is to provide comprehensive reviews on scholarly paper recommendation.
\end{abstract}


\titlepgskip=-15pt

\maketitle

\section{Introduction}
\label{sec:introduction}
\PARstart{R}{ecommendation} has become increasingly important and changed the way of communication between users and web sites. Recommender systems has a large number of applications in many fields such as economy, education, and scientific research, etc~\cite{Kong2018VOPRec,Yu2018PAVE,Trappey2013Intelligent,Son2018academic}. The rapid development of information technology makes the volume of digital information increase quickly~\cite{Xia2017Big,Liu2015Understanding}. Researchers search and filter information such as movies, music, or articles from search engines like Google and Bing by using big data analysis techniques \cite{Fahad2014A,Xia2018Exploring,Liu2018Artificial}. Some researchers share their research findings and publications via digital platforms for free or fee-based access to the Internet for knowledge exchange \cite{Sun2014Leveraging}. The excessive information brings about \emph{information overload} and makes it difficult for researchers to properly judge the relevance of retrieved items for making the right decision \cite{Xia2017Big,Miah2016A}. Recommender systems are introduced in scientific communities to effectively retrieve information \cite{Trappey2013Intelligent,Liu2016CAR,Wang2015A,Aznoli2016Cloud,Zhu2017SEM,Bollen2007Usage,Chua2017Cross}. In academic research, recommender systems can provide papers for researchers and helps them quickly find the papers they need. For instance, for junior researchers with limited publishing experience, recommender systems may recommend new articles and classical articles from related areas for them to broaden their horizons and research interests. On the contrary, for senior researchers with stronger publication records, the recommender systems mainly recommend papers that align to their research interests \cite{Sugiyama2010Scholarly}.
\par Recommending similar scientific articles for researchers is called scientific paper recommendation in scientific community. Paper recommender systems aim to help researchers mitigate information overload and find relevant papers by calculating and ranking publication records, and recommend the top $N$ papers associated to a researcher's research interests or research focus \cite{Wang2011Collaborative}. Nowadays, paper recommender systems have become an indispensable tool in the academic field. Its recommendation algorithms are continuously updated. The accuracy of the recommendation is improving over time. Compared to the traditional keyword-based search technique, recommender systems are more personalized and effective for massive amounts of data \cite{Sun2014Leveraging,Liu2016CAR,Feng2015Personalized,He2016SocoTraveler,dai2018low,sharma2017concept}. The results of keyword-based searching are not always suitable, and the number of items is relatively large \cite{Hassan2017Personalized}. Researchers have to filter the searching results to get the items needed. In the case of different researchers, if they input the same query, they can obtain the same searching results. Because the keyword-based search technique does not consider the users' different interests and purposes. In addition, some researchers don't know how to summarize their requirements, resulting in inputting inappropriate keywords. In comparison, paper recommender systems usually consider researchers' interests, co-author relationship and citation relationship to design the recommendation algorithms and provide the recommendation lists. It should be noted that recommendation results are usually different subjects to researchers' interests. The number of the results can be short and controllable to ensure that the recommender systems is personalised and effective.
\par Since the recommender systems are introduced, many recommendation algorithms have emerged \cite{Xia2016Scientific,Song2017Whose}. The recommendation techniques can be divided into four main categories: Content-Based Filtering (CBF), Collaborative Filtering (CF), Graph-Based method (GB) and Hybrid recommend method. Each method has its own rationale underlying to recommend interesting articles for researchers~\cite{Xia2016Scientific,Sugiyama2011Serendipitous}. CBF mainly considers the users' historical preference and personal library to extract and build users' interest model, which is called user profile \cite{Sun2014Leveraging}. Then CBF extracts keywords from the candidate papers and calculates the similarity of the keywords extracted from user profiles and candidate papers. After ranking the similarity, papers with high similarity will be recommended to users. CF mainly focuses on the actions or ratings on the items of other users whose profiles are similar to the user's called ``neighbour users'' \cite{Pera2014Exploiting}. Users have similar interest in the past, they would probably agree in the future as well. There are many studies about the graph-based method \cite{Zhao2016Paper}. Previous researches construct the graph, in which authors and papers are regarded as nodes. The relationship between papers, the relationship between users and the relationship between users and papers are regarded as edges. Then random walk or other algorithms on the graph are used to compute the relevance between users and papers. For the hybrid method, recommender systems usually use content-based filtering and collaborative filtering method to generate recommendation because the two methods have their advantages and disadvantages respectively. The content-based filtering and collaborative filtering methods complement with each other, the recommender systems with their combination is usually more accurate than the system that only runs a single recommender algorithm. Apart from the three methods above, there are some other paper recommendation techniques: latent factor model \cite{Zhang2013Combining} and the topic regression matrix factorization model \cite{Li2013Scientific}.

\begin{figure}[htb]
\centering
\includegraphics[width=\linewidth]{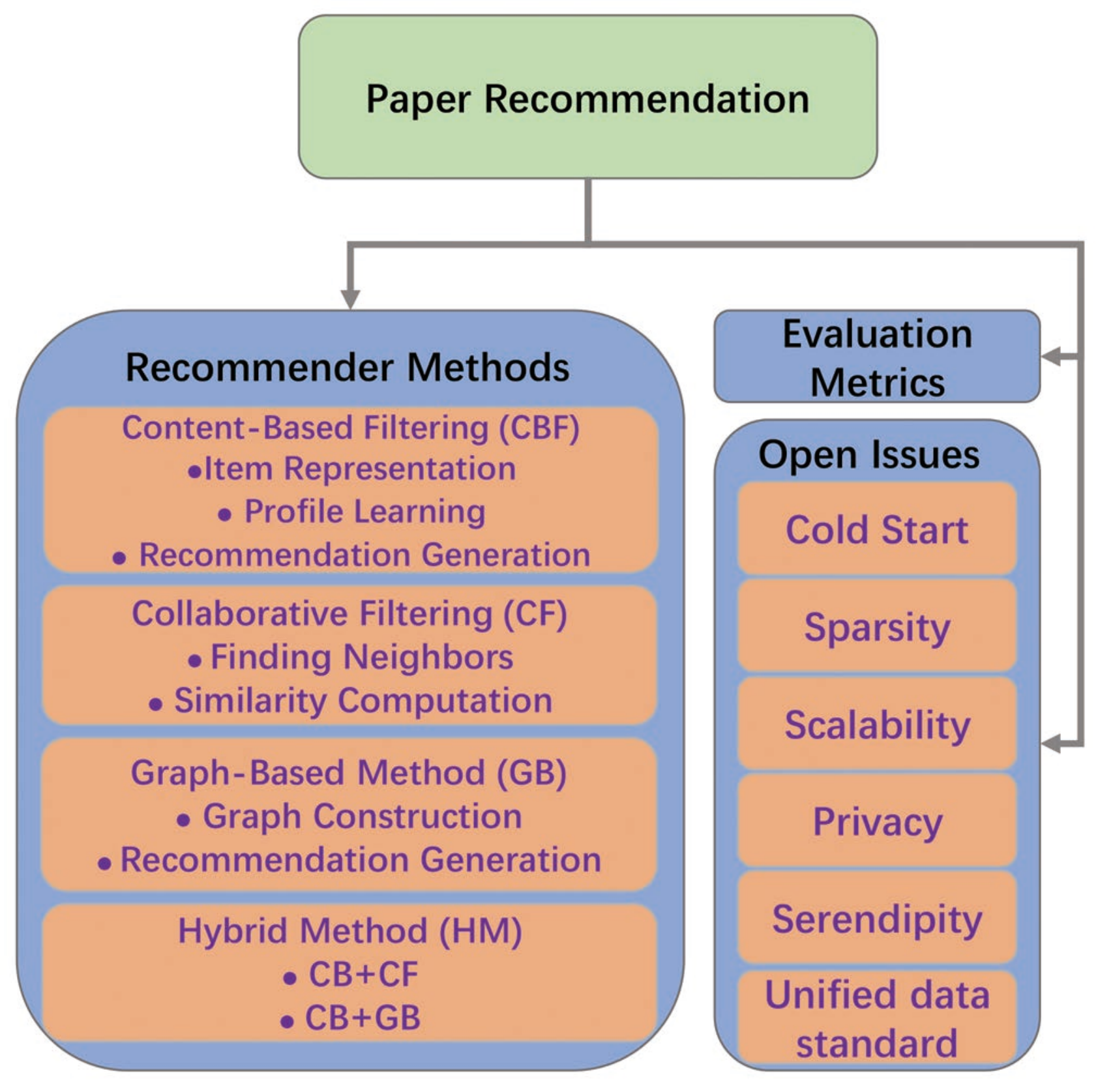}
\caption{Main contents of scientific paper recommendation.}
\label{fig:1}
\end{figure}

\par The main contributions in this survey include:
\begin{enumerate}
\item Classification of commonly used scientific paper recommendation methods.
\item In-depth analysis of the evaluation metrics for paper recommender systems.
\item Summarize problems and challenges in paper recommender systems.
\end{enumerate}

Fig.~\ref{fig:1} shows the main structure of this paper, including recommender methods, evaluation metrics, and open issues. In Section~\ref{sec:2}, we discuss the existing recommendation methods and their research statuses such as Content-Base Filtering, Collaborative Filtering, Graph-Based method, and Hybrid method. The evaluation metrics of the recommender systems are introduced in detail in Section~\ref{sec:3}. Section~\ref{sec:4} summarizes the problems and challenges in the existing paper recommender systems, including cold start, sparsity scalability, privacy, serendipity and unified scholarly data standards. In Section~\ref{sec:4}, we present a summary of this paper.

\section{Paper Recommendation Methods}
\label{sec:2}
In this section, we will overview and discuss the underlying rationale, advantages, disadvantages, and applications of paper recommendation methods.

\subsection{Content-Based Filtering (CBF)}
As a traditional recommendation method, CBF's rationale is simple. The items recommended by CBF method are similar to the items of users' interest \cite{Pazzani2007Content}. Matching information between items and users is the key procedure. In paper recommender systems, items are the papers in the digital library and users are the researchers. In CBF method, a researcher' papers are first collected. Citing the researcher's papers or some other information can be used to build his profile. There are many ways to build researcher profile according our statistics. For example, researcher's preferences and interests can be represented by extracting keywords from researcher's research field. Moreover, paper recommender systems can extract keywords from title, abstract and content of papers to represent these papers. These candidate papers can be retrieved from the digital library. The paper recommender systems then computes the similarity of the keywords between researcher profile and candidate papers, and ranks them. The following  candidate papers with high similarity are recommended to the researcher.
\par
\begin{figure}[htb]
\center{\includegraphics[width=2.8in]{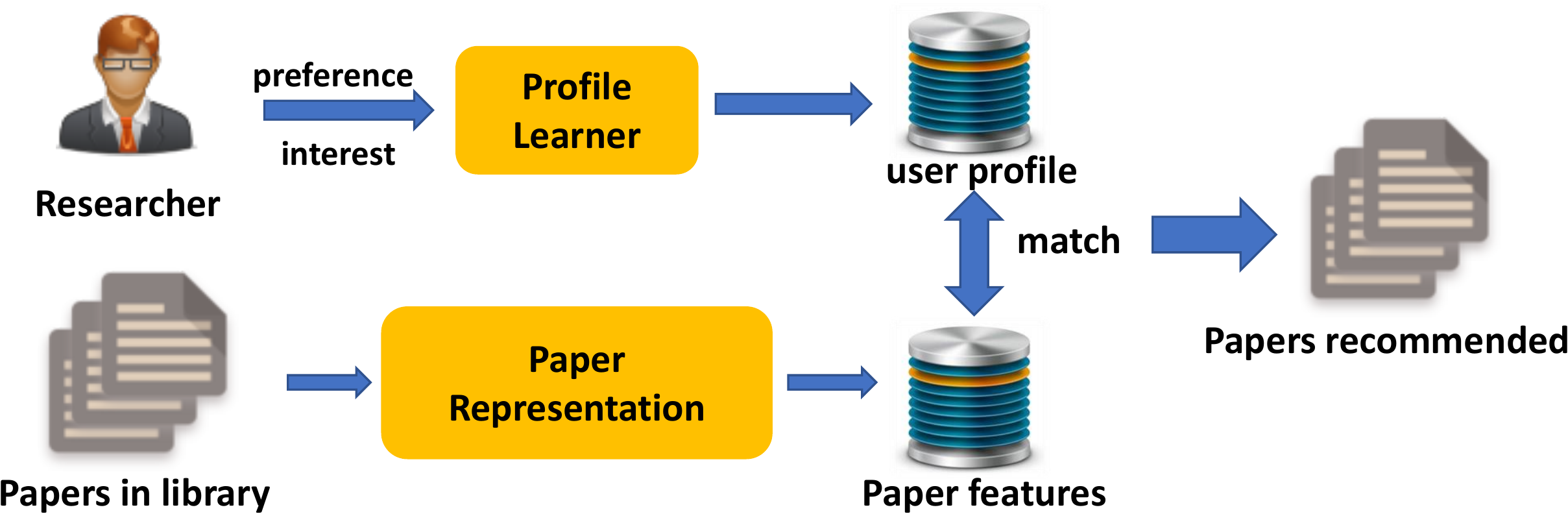}}
\centering
\caption{Content-based system for paper recommendation.}
\label{fig:2}
\end{figure}

 According to the rationale underlying, we can find some advantages of CBF. CBF system extracts paper information and compares them. If the paper is related to researcher's interests, it will be discovered. Furthermore, compared to the keyword-based search engines, CBF usually considers the current researcher's interest, and does not involve other researchers. If researcher' interests change, the recommended result lists will change in the future. Fig.~\ref{fig:2} shows the general structure of the content-based recommender systems. From Fig.~\ref{fig:2} we can see the recommendation progress of the CBF including three main steps: {\slshape Item Representation, Profile Learning and Recommendation Generation}.\\
\par \textit{Item Representation.} In practice, items usually need some special attributes to distinguish each other. These attributes can be divided into two main categories: structured attribute and unstructured attribute. For the structured attribute, the value of attribute is limited and specific. For the unstructured attribute, the value of attribute often means less clear. Because its value is unlimited, which cannot be directly used to analyze. For example, on a dating site, an item is a human being, who has structured attributes such as height, education experience, origin, and unstructured attributes such as a friend's declaration, blog content. Structured data can be used directly, making them easier to manage and use. Unstructured data (such as articles content), on the other hand, are usually required conversion into structured data before being adopted. In paper recommendation area, the whole structures of the papers are similar, but their contents are unlimited, and each author has his/her own writing style. In order to represent all the papers and compute the similarity between them, we need to translate the contents of papers into structured items. Since paper recommender systems are proposed, there are many item representation methods, such as {\upshape TF-IDF} model \cite{Jomsri2010A}, keyphrase extraction model \cite{Caragea2014Citation}, language model and so on.

\par The {\upshape TF-IDF} model (term frequency-inverse document frequency) has been frequently used for information retrieval and text mining \cite{Jomsri2010A}. The TF-IDF value is a statistical measure to evaluate the importance of a word to a document in a collection or corpus. The basic idea of the {\upshape TF-IDF} model is divided into two aspects. On one hand, the more times the keyword {\itshape K} appears in document {\itshape D}, the more important {\itshape K} is for document {\itshape D}. On the other hand, the higher frequency of {\itshape K} appears in different documents, the less importance of {\itshape K} is for distinguishing the documents. The equation is defined as follows \cite{Sugiyama2010Scholarly}:
\begin{equation}
w_{t_{k}}^{Prec}=\frac{tf(t_k,P_{rec})}{\sum_{s=0}^m tf(t_s,P_{rec})}\times\log\frac{N}{df(t_k)}
\end{equation}
where $tf(t_k,P_{rec})$ is the frequency of keyword $t_k$ in paper $p$, $N$ is the paper count in the candidate set, and $df(t_k)$ is frequency of occurrence of keywords $t_k$.

\par CBF uses the {\upshape TF-IDF} model to calculate the feature vectors $f^{Prec}$ of each candidate paper \cite{Sugiyama2010Scholarly,Sugiyama2011Serendipitous}. These vectors can determine how relevant a research paper is to researcher's query \cite{Philip2014Application}. The definition of $f^{Prec}$ is:
\begin{equation}
f^{Prec}=(w_{t1}^{Prec},w_{t2}^{Prec},...,w_{tm}^{Prec})
\end{equation}
where $m$ is the number of distinct terms in the paper, and $t_k(k=1,2...m)$ denotes each term, two vectors for each paper are used as different input queries. This model is popular for CBF recommender systems, many researchers have adopted a modified version in their research. Some researchers realize that when we read a paper, we may be curious about the problem appeared in the paper or the solution to the problem. Thus, they use {\upshape TF-IDF Model, Topic Model and Concept Based Topic Model} to compute the similarity and find the most problem-related papers and solution-related papers to users, satisfying researcher's specific reading purpose separately \cite{jiang2012recommending}. 

\par Apart from the {\upshape TF-IDF} model, a keyphrase (typically constituted by one to three words) extraction model is used to produce a rich description of content of papers \cite{Beel2014The}. The keyphrase list is a short list of keywords that reflects the content of a paper, capturing the main discussed topics and providing a brief summary of its content. In this model, the title, abstract and keywords of a paper are represented by different vectors: $\overrightarrow V_{abstract}$, $\overrightarrow V_{title}$, and $\overrightarrow V_{keywords}$, respectively \cite{Basu2012Technical}. The $\overrightarrow V_{keywords}$ vector is extracted from the ``keyword'' section of the paper. If the paper has not the ``Keyword'' section, the analysis system will regard the most appropriately representative words as the needed keywords \cite{Hong2013UserProfile}.
\\
\par \textit{Profile Learning.} CBF recommender systems assume that researchers have rated ``Like'' or ``Dislike'' on some items and published papers according to individual interests. The objective of this step is to generate the profile model according to researchers' historical actions. Since researcher profile usually includes researcher's research direction, systems can determine whether researcher {\itshape U} likes a new item by this model \cite{Chen2007CONTENT}.

\par It is obvious that researcher profile should rely on the information generated by the researcher. Various methods exist for building user profiles. Previous researchers build user profile with a mixture of topics extracted from the researcher past publications by the \textbf{LDA} algorithm. The vectors $\overrightarrow V_{abstract}$, $\overrightarrow V_{title}$, $\overrightarrow V_{keywords}$ are extracted from the papers of the researcher's historical actions to build profile. The user profile could be updated if researcher publishes or rates new papers in the future.

\par The tag-based information system uses a component named \emph{User Preference Crawler} to crawl the user preference data. The user's profile is constructed by the papers posted by each individual user and a set of tags posted by the users~\cite{Jomsri2010A,Gautam2012An}. Similarly, tags and the set of documents tagged by researchers can be exploited by the key phrase extraction module for building user's profile~\cite{Ferrara2011A}.

\par To facilitate personalization of the recommender systems, junior researchers who published a few papers and senior researchers with many publications could be differentiated \cite{Sugiyama2010Scholarly,Sugiyama2011Serendipitous}. For a paper, the feature vector $f^Prec$  is firstly defined by the TF in {\upshape TF-IDF} model. The definition of $f^Prec$ is the same as equation (2).
\begin{equation}
f^P=(w_{t_1}^P,w_{t_2}^P,...,w_{t_m}^P)
\end{equation}%
where $m$ is the number of the distinct terms in the paper, and $w_{t_k}^P$, $k \in \{1 \dots m\}$, is defined as follows:
\begin{equation}
  w_{t_{k}}^{P}=\frac{tf(t_k,p)}{\sum_{s=1}^m tf(t_s,p)}
\end{equation}
where $tf(t_k,p)$ is the frequency of term $t_k$ in paper $p$.
After getting the feature vectors of papers, the construction of user profile is divided into the two categories: junior researchers and senior researchers. For junior researchers with only one paper $p_1$, the construction of user profile $P_{user}$ will add contribution of the papers cited by $p_1$. For senior researchers with several published papers $p_i (i=1, \dots, n-1)$ in the past, user profile will add contribution of the papers citing $p_i$ and in the reference list of $p_i$. This method makes both senior and junior researchers' profile more specific.

\par All these introduced profile learning methods are relying on researchers' historical records or actions. In some recommender systems, they regard the papers provided by the researcher as input to build user profile \cite{Nascimento2011A,Hanyurwimfura2015An}. After the paper is provided, the needed information for the system will be extracted from the paper's title, introduction, related work, conclusion, references part to determine the user's profile. In addition, to satisfy user's specific reading purpose, the abstract is sometimes divided into two parts : problem description and solution description so that the system could recommend papers from two aspects respectively \cite{jiang2012recommending}.

\par Moreover, there are some other forms to represent user profile. \textit{Docear} is a recommender systems which has the unique feature of utilizing mind maps for information management \cite{beel2013introducing}. The users of \textit{Docear} organize their data in a tree-like data structure, and they build user model from user's mind map collection to match with its digital library. The \textit{Docear} recommender systems have a component named \textit{UserInterface}, which is assigned to contact with users and collect title, author name, domain, topic of the papers. Then the \textit{Docear} recommender systems collect data to store as XML format for user profile, containing domains, topics and keywords \cite{Hong2013UserProfile,Hong2013Personalized,patil5user}.
\\

\par \textit{Recommendation Generation.} The representations of candidate papers and the profiles of researchers are constructed to select the most relevant $N$ items to users. The relevance of researchers' attributes to papers' attributes can be obtained through similarity measure such as cosine similarity. Given two vectors of attributes \textit{A} and \textit{B}, the cosine similarity can be computed as follows \cite{Jomsri2010A}:
\begin{equation}
  Similarity = \cos (\theta) = \frac{A \cdot B}{||A|| \cdot ||B||}
  \label{eq:similarity}
\end{equation}

\par The recommendation of papers uses user profile vectors $P_{user}$ and feature vectors of the candidate papers $F^{Prec}$, which are defined before to compute cosine similarity of $P_{user}$ and $F^{Prec}$ by using equation~\ref{eq:similarity} \cite{Sugiyama2010Scholarly}.

\par Some previous researches not only provide researchers with the most relevant papers, but also provide serendipitous recommendation with the papers from far away fields \cite{Sugiyama2011Serendipitous}. The serendipitous recommendation is helpful for researchers to discover new ideas, approaches or ways of thinking. In serendipitous recommendation researches, researchers construct a basic user profile $P_{user}$ for each researcher $u$ to recommend relevant papers and use $P_{user}$ to construct a another user profile $P_{user}^{srdp}$, then compute cosine similarity between $(P_{user}^{srdp}$ and $F^{Prec})$, $(P_{user}$ and $F^{Prec})$ to generate recommendations. The result of this recommendation has two lists: related papers and unrelated papers.

\par After computing the similarity of user profile and candidate papers, a result list will be generated. The last step of the recommender systems is ranking them in a certain order. The final list top $N$ papers will be recommended to researcher. While ranking the candidate papers, the number of papers citing them is sometimes considered \cite{Pera2011A}.

\par Subsequently, researchers can use this recommender systems to find the paper they are interested in. But there are still some problems in CBF recommender systems. On one hand, CBF does not take the quality such as authoritativeness, style into consideration because its analysis techniques only base on the word analysis. On the other hand, there is the new user problem. If a junior researcher without much research experience uses the system, which perhaps run ineffectively. Because it cannot extract enough information from the user's work, the recommended list may be not reliable \cite{Balabanovi1997Fab}.

\subsection{Collaborative Filtering (CF)}
Like the recommendation techniques of CBF, CF needs to know users' interests, which is especially effective for recommending related papers, even without content-based features \cite{vellino2015recommending}. The basic idea of CF is that if users $A$ and $B$ make ratings on some common items, their interests will be considered similar. If there are some items existing in user $B$'s record but not in user $A$'s, these items can be recommended to user $A$. In other words, CF is the process of recommending items using the opinions of other users \cite{schafer2007collaborative}. The ratings or opinions can be obtained from some social reference management website like CiteULike, or by asking users to fill in a questionnaire \cite{Tang2009A}.

\par
\begin{figure}[htb]
\center{\includegraphics[width=2.8in]{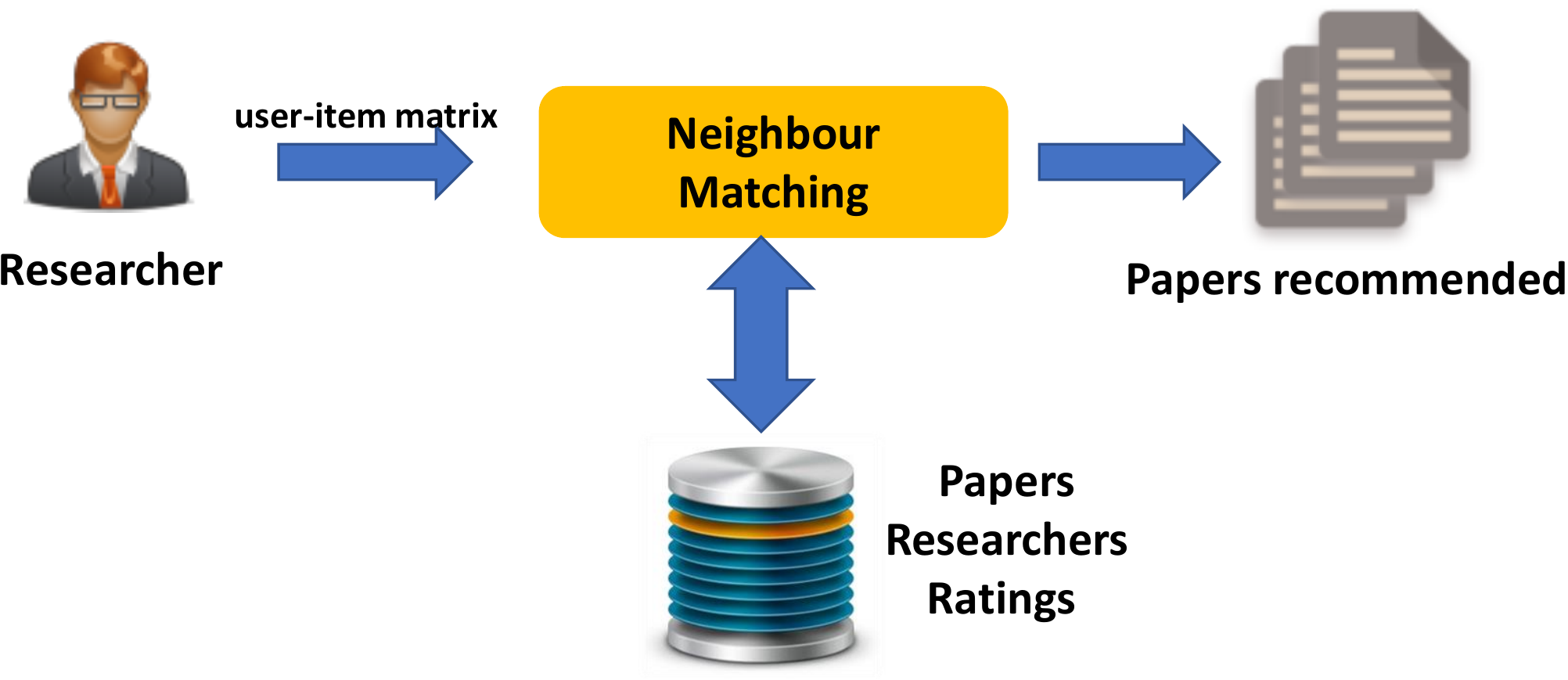}}
\centering
\caption{Collaborative filtering system for paper recommendation.}
\label{fig:3}
\end{figure}

 The collaborative filtering system locates the peer user by considering his rating history and finding the similar user. Then CF uses the neighbourhood to generate the recommendation. The CF Recommender Systems usually need a user-item matrix to represent the users' ratings or comments on items. The ratings can be used to represent users' interests. After constructing the matrix, the system will calculate the similarity between users to find similar users called ``neighbour users'' to recommend items. A user-item matrix is shown in Table~\ref{tab:2}, the elements in the matrix are the users' ratings. In this matrix, the rates are 0 and 1, and the rates can use more numbers to express the different degrees of like or dislike. The general structure of collaborative filtering systems is shown in Fig.~\ref{fig:3}.
\begin{table}[htbp]
  \centering
  \caption{\bf User-Item Matrix}
    \begin{tabular}{|c|c|c|c|c|c|c|}  \hline
      &Item1 & Item2  & Item3   &Item4  & ... & ItemX \\ \hline
 User1 &0 &1 &0 &1 &... &1 \\ \hline
 User2 &1 &1 &0 &0 &... &0 \\ \hline
 User3 &1 &0 &1 &1 &... &0 \\ \hline
 ...&...&...&...&...&...&...\\ \hline
 UserY &1 &0 &0 &1 &... &1 \\ \hline
      \end{tabular}
  \label{tab:2}
\end{table}\\

\par Compared to the Content-Based Filtering method, CF has some different advantages: the content of the recommended paper is not considered, because the recommendation method depends on the ratings made by users and does not consider what kinds of items they belong to. Furthermore, the items recommended to users may not be relevant to the user's current research, because the similarity is measured between the relationships between users.
\par CF mainly contains the two categories of methods \cite{Rana2012Survey}:
\begin{enumerate}
\item[1.] User-based approach: Users are the center in the user-based approach. Recommender systems use the profiles of other similar users to recommend \cite{Xu2018User}. User-based CF finds the nearest neighbours of the users. According to the neighbour's interests, user's interests are predicted \cite{Xu2018User}. Usually, in the user-based systems, users are divided into the several groups, the users in the same group share the same or similar interests on some items. Based on the ratings made by the users in the same group, the recommender systems do recommendation for users.
\item[2.] Item-based approach: Item-based method mainly focus on the relationships between papers rather than users \cite{Joeran2016paper,Valcarce2016Item}. In the item-based approach, there is the assumption that user's interest is continuous or very little change in the future. If users have given some positive ratings on some items, the recommender systems could collect the candidate items by relying on the analysis of  users' rating history. Then the recommender systems will recommend the items by clustering the similar items.\\
\end{enumerate}

\par According to the users' different needs, the above-mentioned recommendation techniques can collect necessary data and recommend papers. The metadata from CiteULike can be used to run CF recommendation algorithm, and it contains many users and their unique tags on papers \cite{Bogers2008Recommending}. The recommendation algorithm is classical and simple: in the user-based filtering, the target user is matched with the collected data to find the neighbours who have similar records. Once the neighbours are found, all the papers of the neighbour's historical preference will be considered as the candidates to recommend to the target users. In the item-based filtering, the system recommends the papers by matching the papers with the target user's historical records.

\par For the user-based CF, the similarity between two users is calculated by the ratings of their common items \cite{Parrasantander2010Improving}. The equation is as follows:
\begin{equation}
Sim(u,n)=\frac{\sum_{i\subset CR_{u,n}}(r_{ui}-\bar{r_u})(r_{ni}-\bar{r_n})}{\sqrt{\sum_{i\subset CR_{u,n}}(r_{ui}-\bar{r_u})^2}\sqrt{\sum_{i\subset CR_{u,n}}(r_{ni}-\bar{r_u})^2}}
\end{equation}
where $r$ is the ratings, $u$ is the target user and $n$ is the neighbour user, $r_{ui}$ stands for the ratings given by user $u$ to item $i$, $\bar{r}$ is the average rating of user $u$ over all his items. $CR_{u,n}$ shows the common set of items between user $u$ and user $n$. The neighbour users' articles are recommended to the target user by ranking the predicted rating for target user $u$. The social relations are usually added to find the proper neighbours. After finding the nearest neighbours, the next step is to predict the target user $u$'s rating for item $i$ \cite{schafer2007collaborative}. The predicted formula is as follows:
\begin{equation}
pred(u,i)=\bar{r_{u}}+\frac{\sum_{n\subset neigh(u)}userSim(u,n)\cdot (r_{ni}-\bar r_{n})}{\sum_{n\subset neigh(u)}userSim(u,n)}
\end{equation}

For a given user-item matrix, the matrix factorization model plays an important role in the collaborative filtering recommender systems \cite{Li2013Scientific}. The matrix factorization model is used to predict the ratings of the candidate papers.

\par The user-based CF algorithms recommend papers in the social tags system \cite{Parrasantander2010Improving}. Researchers summarize the user-based collaborative filtering process as two steps: the first step is to find the neighbours of the target user, the second step is to use the neighbours to rank the items, then recommend top $N$ items for the user \cite{Parrasantander2010Improving}. To improve the quality of recommended result, the two steps are ameliorated \cite{Mishra2014Optimised}. At the finding neighbours step, $BM25-based$ similarity is used to obtain the neighbours of target user \cite{Manning2008Introduction}. At the ranking items step, $Neighbor-weighted$ $Collaborative$ $Filtering$ $(NwCF)$ model is used to calculate the predicted rating. This method improves the original method by considering the number of raters, which is represented as $nbr(i)$. The new predicted rating is computed by:
\begin{equation}
pred'(u,i)=\log_{10}(1+nbr(i)\cdot pred(u,i))
\end{equation}


\par Moreover, scholarly papers are recommended by using the social relationships such as friends, research familiarities \cite{Asabere2015Scholarly}. Besides, the user's profile, group profile and the social relationships between users usually are considered to recommend scholarly papers. For example, a folksonomy based method is used to combine them to recommend, and the method solves the problem that researchers cannot find the relevant scholarly papers in conferences and journals \cite{xia2014folksonomy}.

\par Similar to the user-based approach, the item-based collaborative filtering includes the two steps: similarity computation and prediction generation \cite{sarwar2001item}. At first step, similarity like cosine similarity, thematic similarity of target items $i$ to the set of items rated by target user are used to find the most similar $k$ items {$i_1, i_2, \dots, i_k$} for the candidate item set. In second step, after getting the most similar items, the prediction would be then computed by a weighted average of target user's ratings on these similar items.

\par To guarantee the relevance of the result, an improved item-based collaborative filtering system recommends papers rated by the $connections$ of target user $U$. The recommended papers are not only similar to the target publication $P$ of interest to the target user $U$, but also are popular among the target user $U$'s connections \cite{Pera2014Exploiting}. In this system, researchers first find the target user's connections who exchange and share bibliographic references with target user. Then word correlation factors are used to determine candidate papers $CandidateP$ which are similar to the target paper $P$ from the library of $connections$. Finally, the system recommends the highest ranking scores to target user $U$.

\par From the overview of the CF paper recommendation techniques, we can see the CF is a popular recommendation method. But CF still has some disadvantages because of its natural, and the most obvious shortcoming is the cold start problem. For the new items without ratings, it cannot be recommended until there is someone's rating on it. For the new users with few ratings on any items, his/her rating history is empty, system cannot find a similar neighbourhood until he/she makes enough ratings. To overcome the problems in CF, researchers have thought out some other recommendation techniques, like graph-based method and hybrid method.
\\
\subsection{Graph-Based method (GB)}
\par As the name illustrates, graph-based method mainly focuses the construction of the graph. The graph can be constructed by citation networks, social networks and so on. The researchers and papers are the different nodes of graph. The relationships between researchers, researchers and papers, papers and papers can be considered as the edges between nodes. Then the recommendation system can use an algorithm like random walk on the graph to find the relevant papers for researchers. The advantage of GB is that GB can use information from different sources to recommend. CB, CF just use one or two kinds of information. GB can add social relations, trust relationships between researchers into the recommendation system to make improve the recommendation result.

\par
\begin{figure}[htb]
\center{\includegraphics[width=2.8in]{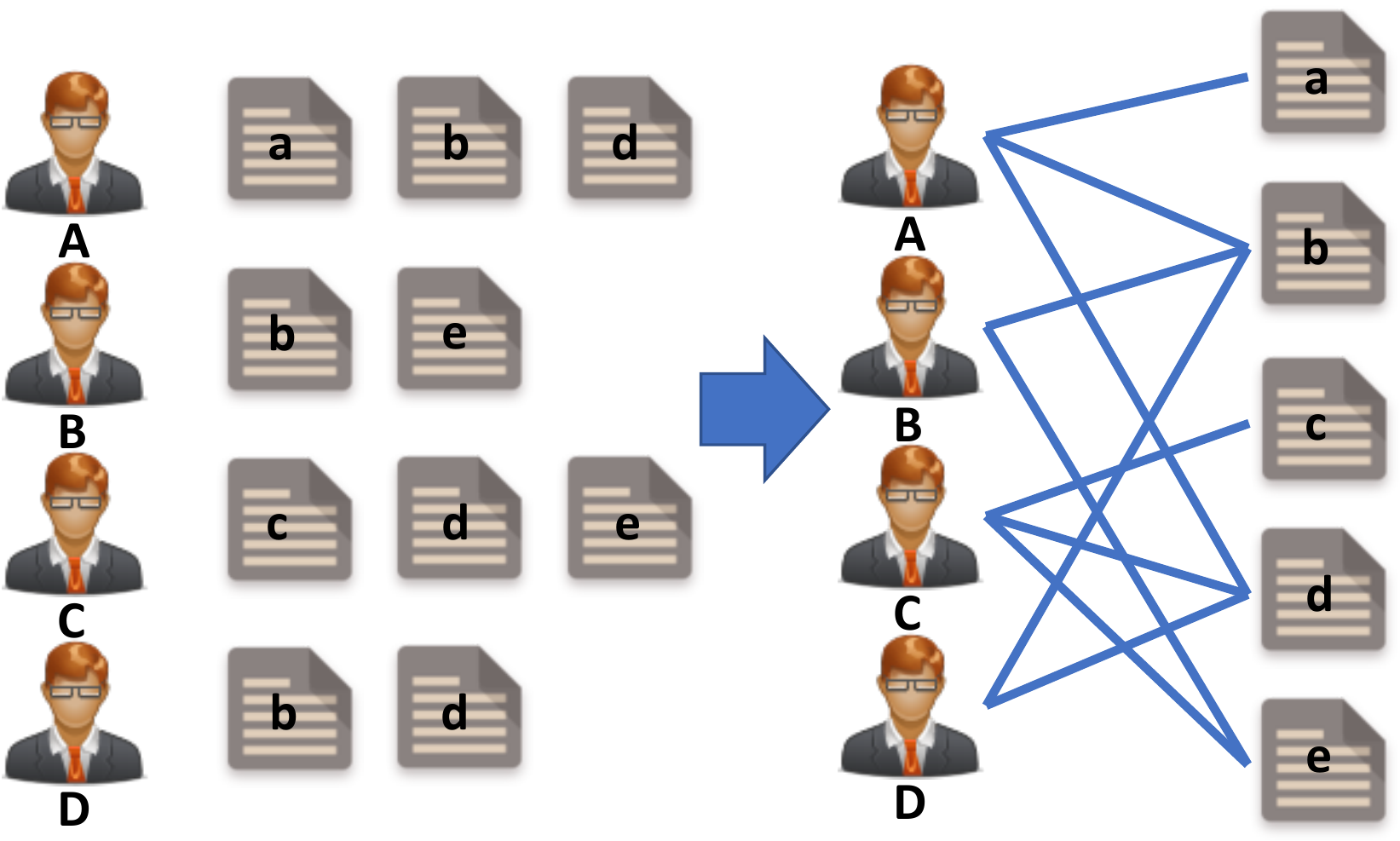}}
\centering
\caption{ A simple graph-based model.}
\label{fig:4}
\end{figure}

 In the graph-based model, we first need to collect data about researchers and papers. Then the system represents them with a heterogeneous graph $G(V,E)$, where $V=V_U \cup V_P$, $V_U$ stands for the researchers in the system and $V_P$ is the set of papers published or referenced by the researchers. For each tuple $(U,P)$, there exists an edge $E(v_u,v_p)$ in the graph, and $v_u\in V_U$, $v_p\in V_P$. There is a simple graph-based model shown in Fig.~\ref{fig:4}. Moreover, in some graph-based recommender systems, there also exist edges like $E(v_u,v_u)$, $E(v_p,v_p)$ which means they consider the relationships between researchers, in addition, they also consider the relationships between papers. In the graph-based model, paper recommendation activity will be translated into the graph search task \cite{huang2002graph}.
\par In Fig.~\ref{fig:4}, $A, B, C, D$ stand for different researchers in the system and $a, b, c, d, e$ represent the papers they have published. The left part is the researcher behaviour data we collect from digital library. The researcher $A$ published paper $a$, paper $b$ and paper $d$, likely researcher $B$ published paper $b$ and paper $e$. We use these researcher behaviour data to build the network in right part. After getting the two-part graph of the researchers and papers, the task of recommender systems can be transformed into calculating the relevance between the unconnected user vertices $v_u$ and paper vertices $v_p$. Many algorithms have been proposed in several papers to recommend relevant papers to researchers \cite{Xia2016Scientific,Zhou2014Authoritative}.

\par The recommendation progress of graph-based recommender systems can be summarized as the two steps: {\slshape Graph Construction and  Recommendation Generation.}
\\
\par {\slshape Graph Construction}. Nowadays many digital materials are used to read and share for people. For academic research, researchers read and search relevant papers from some digital libraries like IEEE Xplore and CiteULike. Researchers can collect data about users and papers from above-mentioned websites to build graph.
\par For example, the relationship between a researcher and a paper means that the researcher is interested in that paper. A matrix $W_{RA}^{n\times m}(i,j)$ is used to indicate whether a researcher $R_{i}$ is interested in the article $A_{j}$ as shown in equation (8). $R$ is a set of $n$ researchers $ {R_1,\cdots,R_n} $. $A$ is a set of $m$ articles $ {A_1,\cdots,A_m} $. The common author relationships are also added into the basic graph \cite{Liu2016CAR,PrabhuEnhancing,Xia2016Scientific}. For the common author relationships between articles, another matrix $W_{AA}^{m\times m}(i,j)$ is introduced to indicate whether two articles $A_i$ and $A_j$ have common author(s) as shown in equation (9).
\begin{equation}
W_{RA}(i,j)=\left\{
\begin{aligned}
1&\quad if\; R_i\; shows\; interest\; in\; A_j \\
0&\quad otherwise
\end{aligned}
\right.
\end{equation}
\begin{equation}
W_{AA}(i,j)=\left\{
\begin{aligned}
1&\quad if\; A_i\; and\; A_j\; have\; common\; author(s)\\
0&\quad otherwise
\end{aligned}
\right.
\end{equation}
After getting the two mentioned matrices, they will be transformed into a graph for further processing. Let $G=(V_R\cup V_A,E_{RA}\cup E_{AA})$, where $E_{RA}\subseteq V_R\times V_A$, and $E_{AA}\subseteq V_A\times V_A$. $V_R$ and $V_A$ are the vertices set of researchers and papers, similarly, $E_{AA}$ represents the set of interest relationships between researchers and papers. $E_{RA}$ represents common author relationships. If $W_{RA}(i,j)$ equals 1, between researcher $i$ and papers $j$ exists an edge in the graph. similarly, if $W_{AA}(i,j)$ equals 1, there is an edge between papers $i$ and article $j$. A hybrid graph with co-author relationships can be built, which is used to generate recommendations.

\par Another heterogeneous graph called ``Bi-Relational Graph (BG)'' can be used to recommend papers \cite{Tian2013Recommending}. BG is similar to the mentioned graph, it also includes researchers and papers. Additionally, BG contains paper similarity subgraph, researcher similarity subgraph, and a bipartite graph connecting researchers and papers.

\par The above heterogeneous graphs contain the two kinds of vertices: researchers and papers. In addition, there is another kind of graph: Citation Graph (Network). Citation graph contains papers and the citation relationship between the papers. The nodes represent the different papers in the citation networks, and the edges stand for the citation relationships between papers. The basic idea in the citation graph is that if two papers have common references or they are cited by one paper, they are considered to be similar \cite{Liu2015Context}. Therefore, the recommendation can be given by analyzing the structure of the citation network.

\par Based on the citation network, a paper $\bar{p}$ can be recommended to user by recommender systems \cite{Zhou2014Authoritative,Gori2007Research}. Let all the papers as $D={p_1,p_2,\cdots,p_n}$ to build a citation graph. $R_{\bar{p}}$ is a subset of $D$, $R_{\bar{p}}$ indicates all the papers cited by $\bar{p}$. Papers in $R_{\bar{p}}$ are related to paper $\bar{p}$. If a paper $p_k$ in $D$ is related to one or more papers in $R_{\bar{p}}$, then paper $p_k$ will be recommended to the user. Based on the similar idea, a method is proposed to recommend papers using citation network and content-based algorithm \cite{Steinert2015Where}. In the weighted heterogeneous graph, researchers replace the author part with the key term graph containing the key terms extracted from each paper using the \textbf{TF-IDF} model. The weight of the citation relationship between the pairwise papers is the cosine similarity of two vectors $p_i$ and $p_j$. The \textbf{TF-IDF} score is the weight of key-term to the paper, and the similarity of two terms is the weight of edges.

\par Moreover, the co-author relationships between authors can be added into the citation network. This graph is called citation-collaborative network. It has the three different kinds of links representing different relationships: citation relationships, collaborative relationships and author-paper relationships \cite{wang2016academic}.

\par The main form of graph construction has been introduced above. There are some other kinds of graphs used to generate relevant papers to the researchers or a given paper from the candidate papers, such as concept map, hub-authority graph \cite{Zhao2016Paper,Paraschiv2016A,Ohta2011Related}.\\

\par {\slshape Recommendation Generation}. The algorithms in the graph-based paper recommender systems usually do not consider the feature of the paper content and the researchers' profile. The reason is that they are not suitable as the nodes of graph for scholarly recommendation. In the graph, researchers and papers represent the two kinds of nodes. The paper recommendation system takes advantage of the information from the graph's structure to find the relevant papers.

\par Random walk with restart algorithm can be used to rank articles \cite{Liu2016CAR,PrabhuEnhancing,Xia2016Scientific,Tian2013Recommending}. The rationale underlying of traditional random walk is that a random walker is used to traverse a graph from one or a series of vertices with the probability ${a}$ of walking to the neighbour vertices of the current vertex and the probability ${1-a}$ of jumping randomly to any vertex in the graph. Each walking gives a probability distribution that indicates the probability that each vertex in the graph is accessed \cite{Fouss2007Random}. This probability distribution is used as the input for the next walk and repeats this process iteratively. When certain preconditions are satisfied, the distribution tends to converge. Random walk with restart method is the improvement on the basis of random walk algorithm. Likely when the walker starts from one node in the graph, it has the probability ${a}$ of moving to the neighbour vertices of current vertex, and the probability ${1-a}$ of returning to the source vertex. The bipartite network uses the random walk with restart algorithm to compute the papers' rankings \cite{Xia2016Scientific}.
\par Moreover, cross-domain recommender systems sometimes use the random walk model. For instance, in a cross-domain recommendation system, they use random walk to find the similar users for the target user \cite{zhenzhen2016cross}. In the study, researchers first use the social relationships to build a network between users. For the target users, the assumption is they tend to accept the recommendation from their friends with similar interests. Therefore, the random walk model is used to get the similar users. Then the systems predict the ratings by the most similar users. Finally, recommendation list is generated. Cross-domain recommender systems aim to build the relationships between the source domain and the target domain, which can alleviate the problems of cold start and sparsity \cite{Niu2016FUIR}, improving the quality of recommendation result.

\par PaperRank is widely used in the recommender systems to calculate the relevance between the papers in citation network \cite{Gori2007Research}. PaperRank is the extension of PageRank model to evaluate the scientific papers, considering the indirect relationships between papers \cite{Du2009PaperRank}. The citation analysis in the previous methods is simple: ISI Journal Impact Factor only averages the citation frequency of the published articles and returns a ranking list of journals \cite{garfield1972citation}. The number of the cited papers is used to rank papers according to the number of direct citation relationships \cite{Garfield1995New}. The rationale underlying of PaperRank algorithm is that it uses papers to replace the pages in PageRank \cite{Haveliwala2003Topic}. Each individual PageRank value can be computed by the following equation:
\begin{equation}
PR(P_i)=\frac{1-d}{N}+d\sum_{i\neq j}\frac{PR(P_i)\cdot l(P_i,P_j)}{L(P_j)}
\end{equation}
where $P_1,P_2,\cdots ,P_N$ are the $N$ papers in the citation network, $PR(P_i)$ is the PageRank value of paper $P_i$ (ie. ranking score of the paper), $L(P_i)$ is the number of the paper $P_i$'s reference papers, $d$ is the damping coefficient, $l(P_i,P_j)$ is the function of whether paper $P_i$ cited paper $P_j$. if $P_j$ is cited by $P_i$ then $l(P_i,P_j)$ equals 1, otherwise $l(P_i,P_j)$ equals 0. Using this method, the importance of the individual papers can be expressed.

\par Using the structure of the graph to recommend papers is a novel method. The GB mainly uses the relationships between the nodes.
\subsection{Hybrid Method (HM)}
\par
\begin{figure}[htb]
\center{\includegraphics[width=2.8in]{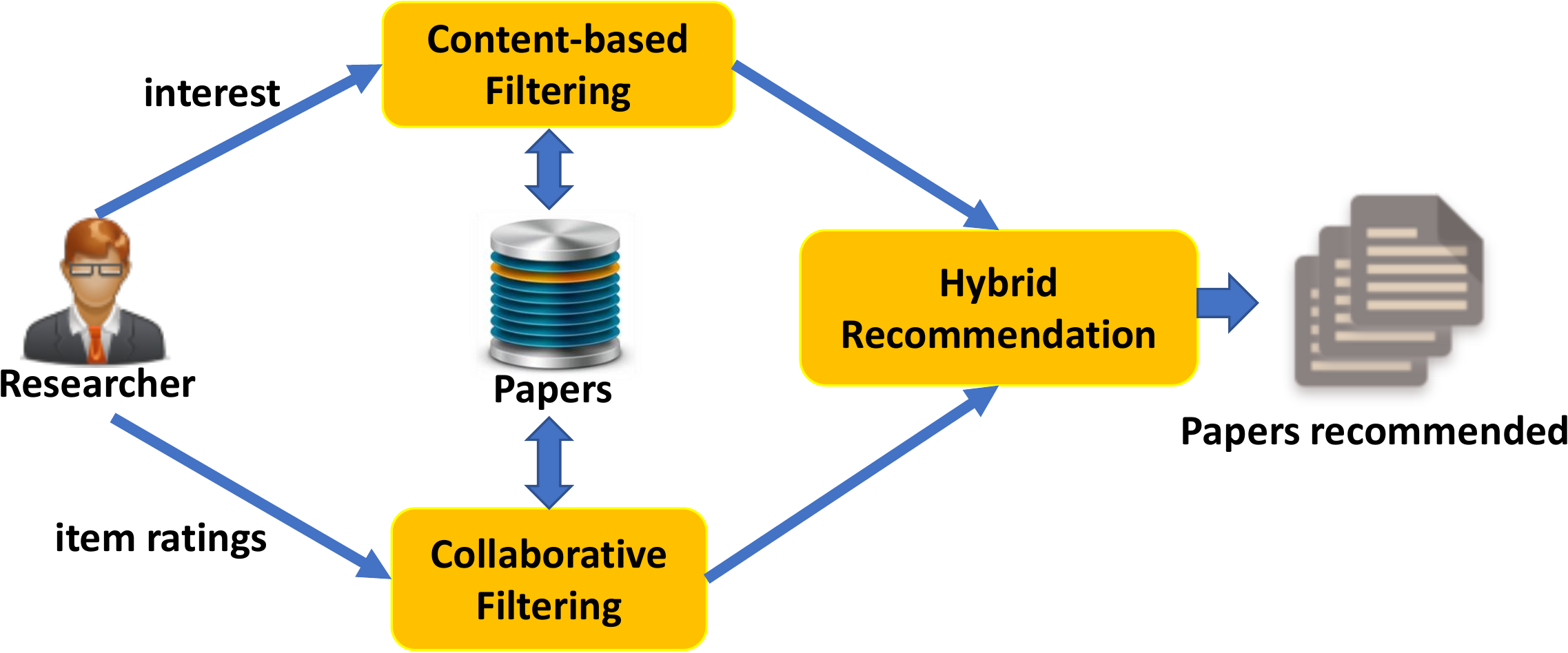}}
\centering
\caption{ A hybrid paper recommendation system.}
\label{fig:5}
\end{figure}

 To improve the accuracy of the recommendation results and obtain the better performance, some scientific paper recommender systems combine the two or more recommendation techniques to recommend the personalized papers to the researchers \cite{Christidis2016Research}. The obvious advantage of HM is that HM can use the combination of different recommendation techniques and the information from many sources. In this section, we introduce some hybrid recommendation techniques. Fig.~\ref{fig:5} shows a hybrid paper recommender systems using the combination of content-based and the collaborative filtering methods.

\par {\slshape Content-based+Collaborative Filtering}. Both the content-based recommendation method and the collaborative filtering method have their own advantages and disadvantages. Some prior studies tried to combine the two methods with different forms to make better paper recommendation and overcome their shortcomings such as first-rater and sparsity problem \cite{Sun2014Leveraging,Winoto2012Contexts,Sugiyama2013Exploiting}.

\par There is a hybrid recommender systems using the content-based techniques and the collaborative filtering techniques. The content-based techniques build researcher's profile by capturing previous research interests embodied in their past publications. The collaborative filtering techniques aim to discover the potential citation papers \cite{Sugiyama2013Exploiting,Herlocker1999An}. The process of recommending papers includes three steps. First, researchers need to build the user profile $P_{user}$ from his/her published papers by using the \textbf{TF} schemes. And they compute feature vectors $F^{P_{i} (i=1,\cdots,t)}$ for each candidate papers by \textbf{TF-IDF} scheme. They find \textit{N} papers with the highest cosine similarity scores. Second, for these papers, CF algorithm operates on the paper-citation matrix based on an idea that similar papers have similar citations to find the potential papers. Pearson correlation coefficient between citation vectors to the target paper is used to measure the similarity. Papers with highest similarity with target paper will be formed as $neighbourhood$ papers.  Finally, the cosine similarity of the content will be computed \cite{Sun2014Leveraging,Zhang2008A,Gipp2009Scienstein,Amami2017A}. By combining the two methods, this system yields superior performance over the classic recommender systems.

\par Base on the traditional recommendation techniques, some modified algorithms have emerged such as {\slshape CBF-Separated, CF-CBF Separated and CBF-CF Parallel} algorithms \cite{Hammou2018APRA}. The \textit{CBF-Separated} algorithm is built upon the pure CBF algorithm. It recommends the related paper lists not only for the target paper itself but also for its references. These recommendation lists are merged into one single list for the researcher. In the \textit{CF-CBF Separated} algorithm, CF method is first used to generate a list of candidate papers to recommend. CBF then is used to give further recommendations based on the list generated from CF. \textit{CBF-CF Parallel} algorithm runs both CF and CBF methods in parallel and generates recommendation lists by combining the result lists from the two methods through an ordering function to make sure the right order of the result list. All these hybrid algorithms are proved to be better than the single recommendation technique.

\par In addition, there are some special hybrid methods such as collaborative filtering with latent factor model, probabilistic topic model \cite{Wang2011Collaborative}, spreading activation model \cite{Zhang2011A}, EIHI algorithm \cite{Dhanda2016Recommender}, FP-growth algorithm \cite{Igbe2016Incorporating}, etc. The performance of these hybrid methods are better than the baseline methods.

\par The latent factor models are used for the collaborative filtering to recommend papers according to other users' historical records or interests, which are similar to the target user's interest. This model is used to recommend known papers \cite{Wang2011Collaborative}. Spread activation model is used in content-based method and user-based collaborative filtering method to find users who have similar interest with the target user \cite{Zhang2011A}. EIHI algorithm is designed to work in the dynamic datasets like the increasing digital library of the published papers \cite{Dhanda2016Recommender}. Embedding EIHI into the content-based paper recommendation system can make the results of recommendation up-to-date and personalized. To guarantee the recommended papers' content and quality, CBF is often used to retrieve all the possible papers in the library. A multi-criteria collaborative filtering is used to find the papers with high quality from the result of CBF \cite{Naak2009A}.

\par {\slshape Content-based+Graph-based}. The combination of content-based method and graph-based method can perform better than the classic recommendation methods. Because the content-based method can gain the user profile from the content of papers that users are interested in. The graph-based method can use the citation network or the bipartite graph to find more potential candidate papers from the structure of the graph.
\par The content-based techniques with citation network have the ability to recommend the most relevant papers from the digital library \cite{Beel2017Mr}. The bipartite graph includes the two layers: papers' layer connects papers with citation relationships. The researchers' layer connects researchers with their social relationships. Specially, to make the recommendation more accuracy, a novel hybrid article recommendation method integrating the social information are proposed \cite{Wang2018HAR}. The recommendation method includes the three types of relationships: (1) For researchers $A$ and $B$, the basic trust is that researcher $A$ and researcher $B$ have overlapped in their library. (2) The value of researcher $B$ will be increased if the researcher $B$ is the author of some papers in researcher $A$'s library. (3) is that researcher $A$ trusts in researcher $B$'s knowledge in special topic. Candidate papers (CP) are from the structure of the bipartite graph. The recommendation system selects CP from the libraries of the current researchers. While building researchers' profile, the junior researchers and senior researchers are distinguished. Both the senior and junior researchers' interests are represented by the feature vectors through the \textbf{TF-IDF} model to analysis the content of the papers. The ranking of the CP will consider the similarity between CP' feature vectors, the researchers' profile, the value of trust between the CP's owner, current researcher, the citation count of the CP, and the reputation of authors.

\par Apart from being combined together, the recommendation methods can be used separately. The content-based method using TF-IDF model gets the feature vectors from the candidate papers. The similarity is gained by computing the cosine similarity of candidate papers and the papers in the target user's record. The graph-based method using the classic citation network runs the BP algorithm and other algorithms to obtain the user's preference and recommend top $N$ papers to the user. The hybrid approach uses the result lists from the two mentioned methods and gives them different weight. Let the $f_{content}$ is the result of the content-based method, $f_{graph}$ is the result of graph-based method, the hybrid result $f_{hyrid}$ is computed as follows:
\begin{equation}
f_{hybrid}=w \times f_{content}+(1-w) \times f_{graph}
\end{equation}
where $w$ and $(1-w)$ represent the weights of the two methods. The combination can solve the over specialization problem and the new item problem of the classic methods.

\par We can see that HM has many different combinations and it uses many techniques. The aim of HM is to improve the quality of recommendation results by using the pros of different techniques while overcoming the cons. The most important problem of HM is the effective Combination of techniques .

\subsection{Others}
\par Apart from the paper recommendation methods mentioned before, researchers invent some other paper recommendation techniques such as modified latent factor model \cite{Zhang2013Combining}, hash map \cite{Honarvar2015A}, bibliographic coupling \cite{Habib2017Paper}, etc. In this section, some novel paper recommendation techniques will be introduced.

\par As shown in the hybrid recommendation techniques, the latent factor model is used to represent the content of papers. The model uses the user-item matrix, papers' content (title, abstract), attributes (author, publish year), and social network as input. The model then uses a modified topic modelling involving the content and attributes to represent users and papers. The matrix factorization method is used to predict according to the user vector $V_u$, the paper vector $V_p$ with the results of topic modelling, and the user-item matrix \cite{Koren2009Matrix}. The paper recommendation result list is from the papers with the highest predict ratings.

\par It is a fact that in the research paper recommendation domain, the number of researchers is much less than the number of papers. While building the citation matrix or the user-item matrix, there are many empty elements. To avoid this problem, the non-sparse matrices are used to represent citation graph of papers, and local sensitive hashing (LSH) constructed a representation of citations in a paper \cite{Honarvar2015A}. An example of traditional and non-sparse matrix representation of citation network is shown in Table~\ref{tab:3} and Table~\ref{tab:4}

\begin{table}[htbp]
  \centering
  \caption{\bf A matrix represents papers with citations.}
    \begin{tabular}{|c|c|c|c|c|c|c|}  \hline
   &P1 & P2  & P3   & P4  \\ \hline
 C1 &0 &1 &0 &1  \\ \hline
 C2 &1 &1 &0 &0  \\ \hline
 C3 &1 &0 &1 &1  \\ \hline
 C4 &1 &0 &0 &0  \\ \hline
 C5 &0 &1 &1 &0  \\ \hline
      \end{tabular}
  \label{tab:3}
\end{table}
\begin{table}[htbp]
  \centering
  \caption{\bf A non-sparse matrix of Table ~\ref{tab:3}.}
    \begin{tabular}{|r|r|r|r|r|}  \hline
      P1 & P2  & P3   & P4  \\ \hline
      C2 & C1  &C3  &C1     \\ \hline
      C3 & C2  &C5  &C3     \\ \hline
      C4 & C5  \\ \cline{1-2}
      \end{tabular}
  \label{tab:4}
\end{table}

In the Table ~\ref{tab:3}, the columns of the matrix represent the citing papers, and the rows represent the cited papers. The sparsity comes from the fact that the matrix should include all the cited papers. For each cited paper, there is a matrix row, but each citing paper in the matrix only cites a part of the cited papers. The non-sparse matrix is shown in Table~\ref{tab:4}, Table~\ref{tab:3} and Table~\ref{tab:4} represent the same citation relationship: P1 cites C2, C3 and C4; P2 cites C1, C2 and C5 $\cdots$. On each row of the non-sparse matrix, there is a hash function, the similarity depended on these functions.

\par Moreover, there exists some other techniques applied in scientific paper recommender systems to provide service to researchers. To improve the performance of the CBF method, CBF is used as the pre-processing step \cite{Ravi2017Cross}, then Long Short-Term Memory (LSTM) method learns a semantic representation of the candidate papers \cite{Hochreiter1997Long}. Finally, the top \textit{N} papers in result list with high content and semantic similarity to input paper. To help junior researchers read more classic papers online, the two principles (download persistence and citation approaching) are proposed to determine whether a paper is a classic paper, which will be recommended to the junior researchers \cite{Wang2010Claper}. A Citation Authority Diffusion (CAD) methodology is proposed to identify the key papers \cite{Chen2012Novelty}. Techniques like Multi-Criteria Decision Aiding \cite{Matsatsinis2007A,Manouselis2007Analysis}, Bibliographic Coupling \cite{Habib2017Paper}, Belief Propagation (BP) \cite{Naak2009A,Ha2014Recommendation}, Deep learning \cite{Hassan2017Personalized}, Canonical Correlations Analysis (CCA) \cite{Gupta2017Scientific}, Singular Value Decomposition (SVD) \cite{Ha2015On} appear in some researches to recommend papers.

\subsection{Comparisons of common techniques}
\par Now we have introduced all the recommendation techniques existing in the papers we collected. There is a comparison table of the common recommendation techniques Content-Based Filtering, Collaborative Filtering and Graph-Based method.
\begin{table}[htbp]
  \centering
  \caption{\bf Comparisons of common techniques}
    \begin{tabular}{|p{1.7cm}|p{2.5cm}|p{2.5cm}|}  \hline
        Technique               &     Advantage    &     Disadvantage\\ \hline
        Content-Based Filtering &  $\bullet$  Each paper can be discovered to compute similarity  &  $\bullet$ Only consider the word relevance quality is  uncertain \\
                                &  $\bullet$  Results are related to users' personal preferences  &   $\bullet$ New user problem  \\ \hline
        Collaborative Filtering &  $\bullet$  Recommendation results may be serendipitous         &   $\bullet$ Cold start problem \\
                                &  $\bullet$  The quality of results can be guaranteed            &   $\bullet$ Sparsity problem   \\ \hline
        Graph-Based method      &  $\bullet$  Considers different source to recommend             &   $\bullet$ Does not consider papers' content and users' interests  \\ \hline
      \end{tabular}
  \label{tab:5}
\end{table}
Table~\ref{tab:5} shows the advantages and disadvantages of CBF, CF and GM. Each recommendation technique can overcome the disadvantages of other techniques. CF can overcome the quality problem of recommendation results, but it still has cold start and other disadvantages. To combine the advantages and avoid disadvantages of these techniques, here comes the hybrid method. The hybrid method uses CBF and CF to make the recommendation system more efficient, in addition, CBF and GB are used to recommend papers.
\section{Evaluation Methods}
\label{sec:3}
As described in Section $\amalg$, there are so many techniques used in the scientific paper recommender systems. All of them can provide researchers some papers, which are related to the input query or researchers' profile. The more recommendation techniques are proposed, the more important their evaluation methods are \cite{Langer2013A,Beel2015A}. The type of evaluation metrics depends on the type of recommendation techniques \cite{Isinkaye2015Recommendation}. The result of the evaluation methods determines whether the technique applied in recommendation system is effective. In this section, we will review the evaluation methods in the recommender systems. Some most frequently used metrics are shown in Table~\ref{tab:6}.

\begin{table}[htbp]
  \centering
  \caption{\bf Classification of evaluation methods}
    \begin{tabular}{|p{1.0cm}|p{1.1cm}|p{0.7cm}|p{0.8cm}|p{0.6cm}|p{0.6cm}|p{0.3cm}|}  \hline
        &Precision & Recall  & NDCG  & MRR   & MAP    &F1   \\ \hline
 Number &20           &20       &14    &9     &5      &4 \\ \hline
      \end{tabular}
  \label{tab:6}
\end{table}
From Table~\ref{tab:6}, we can see that $Precision$ and $Recall$ are the most frequently used evaluation methods in the papers we reviewed. Many paper recommender systems used more than one metrics to evaluate their recommendation techniques. Apart from the metrics in Table~\ref{tab:6}, there are some other less used metrics in the reviewed paper, like $RMSE$, $UCOV$ and $MAE$, all of them will be introduced at the end of this section.

\par $Precision$ : It is used to measure the accuracy of the recommender systems recommending relevant papers to the researchers, the equation is:
\begin{displaymath}
Precision = \frac{Relevant\;papers}{Total\;recommended\;papers}
\end{displaymath}
A bigger value of this fraction indicates the more accurate recommendation that recommendation system made. To reduce the statistics complexity of all papers in the recommendation result, there is a modified version $P@N$ \cite{Ha2015On}.

\par $Recall$: it quantifies the fraction of relevant papers in the whole set of papers that are in the recommendation result list. Its equation is as follows:
\begin{displaymath}
Recall=\frac{Relevant\;papers}{Total\;relevant\;papers}
\end{displaymath}
the denominator in this equation is fixed because the number of the all relevant papers in the library is fixed. The value of equation depends on the rank algorithm of the recommendation system. The bigger value means that recommendation system has ability to rank the most relevant papers at the top of the result list. Similar to $Precision$, $Recall$, modified version $Recall@m$ is the number of relevant papers in the top m of ranking list.

\par $F-measure$: it considers that $Precision$ and $Recall$ could contradict each other \cite{Xia2016Scientific}. From their equations, we can see that  when the number of recommended list becomes bigger, then $Recall$ may grow while $Precision$ may drop. $F-measure$ considers them together and gives a weighted harmonic average of $Precision$ and $Recall$ :
\begin{displaymath}
F=\frac{(\alpha^2+1)(Precision \times Recall)}{\alpha^2(Precision+Recall)}
\end{displaymath}
Due to the fact that $Precision$ and $Recall$ are in the range of $[0,1]$, a high $F$ value means that the paper recommendation system is more effective.

\par $NDCG\;(Normalized\;Discounted\;Cumulative\;Gain)$: it is used to evaluate the quality of a given sorted recommended list \cite{Zhang2011A}. In order to compute the $NDCG$ of the $j$th paper in the result, the average $DCG$ will be computed at first:
\begin{displaymath}
DCG=\frac{1}{|U|}\sum_{u=1}^{|U|}\sum_{j=1}^J\frac{g_{uj}}{max(1,\log_b(j))}
\end{displaymath}
where $U$ is the set of users who participate in this paper recommendation system, $|U|$ is the number of users in $U$, $J$ is the number of papers recommended to users, $j$ is the position of recommended paper in the recommended list, $b$ is a constant value, and $g_{uj}$ represents the ``gain'' that user gets from paper $j$. Base on $DCG$ the definition of $NDCG$ is as follows:
\begin{displaymath}
NDCG=\frac{DCG}{\max{DCG}}
\end{displaymath}
the gain that user gets from recommended papers depends on the quality of recommended papers. If the user thinks that paper is very relevant to his/her research, the gain is high, otherwise the gain is $0$. It is desirable that the most relevant papers appear at the top of the recommended list.

\par $MAP\;(Mean\;Average\;Precision)$: it is invented to solve the single point value limitation from the three introduced metrics: $Precision$,  $Recall$ and $F-measure$. It would be calculated by averaging over all the average precision (AP) of the recommended result for each user \cite{Parra2010Improving}. The definition of $AP$ is:
\begin{displaymath}
AP=\frac{1}{m}\sum_{k=1}^NP(R_k)
\end{displaymath}
where for a user $u$, $m$ is the number of relevant papers to $u$, $N$ is the whole number of the papers in recommended list, $P(R_k)$ represents the precision of retrieved results from the top result until get to paper $k$ \cite{Sun2014Leveraging}.

\par$MAP$: it gives an average of each user's $AP$ value:
\begin{displaymath}
  MAP=\frac{1}{U}\sum_{k=1}^UAP(k)
\end{displaymath}
where $U$ is the whole number of the users involved in this recommendation system.

\par $MRR\;(Mean Reciprocal\;Rank)$: similar to $NDCG$, this metric is used to determine the quality of the sorted recommended paper lists. It only concerns about the ranking of the relevant papers in the recommended list and gives an average over all relevant papers. The definition is:
\begin{displaymath}
MRR=\frac{1}{N}\sum_{i=1}^N\frac{1}{rank_i}
\end{displaymath}
where $N$ represents the number of target papers and $rank_i$ is the rank of $i$th target paper.

\par These metrics can effectively evaluate the various paper recommendation algorithms of the recommender systems from different aspects. These metrics are popular with the researchers of the recommender systems. A good recommendation system must get high score on these metrics. Additionally, there are some evaluation metrics which are rarely applied to the system.

\par $RMSE\;(Root\;Mean\;Square\;Error)$: it is to identify the difference between rating values and predicted values generated from recommender systems \cite{Joeran2016paper}. The true values in the training/testing set can be computed as follows:
\begin{displaymath}
RMSE=\sqrt{\frac{1}{N}(r_{ij}-\hat{r_{ij}})^2}
\end{displaymath}
where $r_{ij}$ is the true rating value, $\hat{r_{ij}}$ is the predicted rating value and $N$ is the number of ratings in the test set. The lower the $RMSE$ is, the stronger the predictive power of the recommendation system.

\par $MAE\;(Mean\;Absolute\;Error)$: similar to $RMSE$, this metric is used to evaluate the accuracy of prediction made by recommendation algorithms \cite{Naak2009A}, it can be calculate by the following equation:
\begin{displaymath}
MAE=\frac{1}{n}\sum_{i=1}^n|f_i-y_i|
\end{displaymath}
where $n$ is the number of predictions, $f_i$ is the prediction rating of paper $i$ and $y_i$ is the true value. The lower the $MAE$, the more accuracy the recommendation system predicts ratings is.

\par $UCOV\;(User\;Coverage)$: because of the nature of recommendation algorithms, there usually exists some users who cannot get useful information from the recommendation system, they cannot get relevant papers from the system. The equation is simple:
\begin{displaymath}
UCOV=\frac{U'}{U}
\end{displaymath}
where $U'$ is the number of users who get relevant recommendations and $U$ is the number of all the users in the system \cite{Parra2010Improving}. Thus, a good recommendation system can be useful for most users not only for a special kind of users in the system.

\section{Open issues and Challenges}
\label{sec:4}
\par In previous sections, we have discussed the recommendation methods and evaluation methods of the scientific paper recommender systems. Although the mentioned paper recommender systems can provide researchers some useful papers by running their own recommendation algorithms, they still have some problems need to be solved and improved. In this section, we discuss some open issues and challenges of the existing paper recommender systems, including Cold Start, Sparsity, Scalability, Privacy, Serendipity and Unified data standards.

\subsection{Cold\;Start}
\par Cold start problem is an important issue of new papers and new users in recommender systems \cite{Kumar2016Approaches}. On one hand, if recommender systems are based on pure collaborative filtering method, they will suffer challenges from both new papers and new users \cite{Schein2002Methods}. For a new user who has no research experience or rarely rates on the papers he/she reads from the digital library, user-based CF cannot find the similar users or neighbours for new user accurately. For a new paper newly published in the digital library, few researchers have read and rated it. The new paper cannot be recognized easily from so many papers and recommended to the right researchers. On the other hand, in the content-based recommender systems, researchers use content analysis to represent all the papers and compute the similarity between papers and user profile, overcoming the new paper problem. But CBF needs to analyze the researchers' historical records containing the papers that a user expresses interest in. If CBF cannot extract enough useful information to build user profile, the result of recommender systems is not reliable.

\subsection{Sparsity}
\par In most recommender systems, there is an assumption that the number of users is bigger than the number of papers or equivalent to the number of papers in digital library. The recommendation algorithms can run effectively. However, the fact is that the number of users is less than the papers, and even the most popular papers may have a few ratings. While building the user-item rating matrix in the collaborative filtering method, researchers find that rating matrix is very sparse, there are too few ratings and too few correlations between users \cite{Luo2017An}. If most of the papers have few ratings and each user only rates on a few papers, it is hard to find the similar neighbours for users. It is one of the most obvious disadvantages of collaborative filtering based recommender systems.

\subsection{Scalability}
\par The definition of scalability in recommender systems is whether the system has the ability to work effectively in numerous environments where there are so many users and products. Nowadays the datasets of the digital library are very large, and the states of papers in it are changing with time \cite{Kumar2016Approaches}. There are many papers and users added into dataset every day. It is challenging for recommender systems to deal with these large and dynamic datasets. Traditional recommendation methods like CBF and CF usually dealt with the static dataset, new learning algorithms like EIHI can handle the dynamic datasets \cite{Dhanda2016Recommender}. It is desirable that each recommender systems can overcome the scalability problem.

\subsection{Privacy}
\par Paper recommender systems aim to provide the personalized paper recommendation to the users by taking advantage of the users' personal information. With the recommendation system widely used in academic area to solve the information overload problem \cite{Lam2006Do}, most personalized recommender systems collected as much users' information as possible. Because the information collected by the system usually includes sensitive information that users wish to keep private, users may have a negative impression if the system knows too much about them \cite{Kumar2016Approaches}. It is an important topic that how to improve the recommendation algorithm by using the limited data fully, carefully and meticulously. To resolve this problem, some secure recommender systems are proposed to protect users' private information \cite{Chen2007CONTENT,Lam2006Do}.

\subsection{Serendipity}
\par The traditional paper recommender systems usually provide users with the papers relevant to his/her interests or researches \cite{Sugiyama2013Exploiting}. In fact, the irrelative papers perhaps have some advantages for users. For example, junior researchers need to read various kinds of papers to broaden their research range and find the most interesting one. Senior researchers need to find new knowledge from other areas to enrich their own studies \cite{Sugiyama2011Serendipitous}. The serendipitous recommendation for users sometimes can be useful, but if the result of the recommendation system only has serendipitous papers and does not have related papers, user may think the system is not reliable. Collaborative filtering method based system has the ability to provide serendipitous results because the recommendation algorithm does not consider the content of the paper only use the ``neighbours'' to recommend items.

\subsection{Unified Scholarly Data Standards}
\par Part of big scholarly data comes from different academic platforms such as Google Scholar, Web of Science, and Digital Bibliography \& Library Project (DBLP). The other part comes from online data sets such as Microsoft Academic Maps and American Physical Society (APS). These data have their own characters. For example, the DBLP data set does not contain citation relationship, and the APS data set provides a list of citation relationship between the papers. These different data types bring a huge challenge to the construction of the paper recommender systems. In the paper recommendation systems, unifying big scholarly data standards is a challenging task.
\section{Conclusion}
\label{sec:5}
\par Recommender systems play an important role in information retrieval and filtering. This paper gives a survey of scientific paper recommendation systems for academic area. First,
we classify the scientific paper recommender systems into four groups according to their recommendation techniques: content-based filtering, collaborative filtering, graph-based method and Hybrid method. According to our analysis, we find the content-based and hybrid methods are the most often used techniques in paper recommender systems. For each technique, we investigate the underlying rationale, advantages, disadvantages and applications. Second, the evaluation metrics are introduced to evaluate the performance of paper recommender systems: Precision, Recall, F-measure, NDCG, MAP, MRR, MAE and UCOV. Finally, this paper discusses the open issues and challenges that need to be solved in the future, including cold start, sparsity, scalability, privacy, serendipity, and unified scholarly data standards.

\EOD

\end{document}